\documentclass[a4paper,11pt]{article}

\usepackage{jcappub}
\usepackage{slashed}
\usepackage{wrapfig}
\usepackage{bm}
\usepackage{latexsym,amssymb,amsmath,float,url,mathrsfs}
\usepackage{latexsym}
\usepackage{graphicx}
\usepackage{epstopdf}
\usepackage{amsfonts}
\usepackage{amsmath}
\usepackage{amssymb}
\usepackage{comment}
\usepackage{subfigure}
\usepackage{natbib}
\usepackage{hyperref}
\usepackage{pifont}
\usepackage{blindtext}
\usepackage{adjustbox}
\usepackage{multirow}
\usepackage{tabularx}
\usepackage[table]{xcolor}
\usepackage{orcidlink}
\usepackage{tikz}
\usetikzlibrary{tikzmark}
\usepackage{amssymb}
\usepackage{bbold}
\usepackage{orcidlink}

\usepackage{blkarray}
\usepackage{graphicx}
\usepackage{amsfonts}
\usepackage{soul}
\usepackage{amssymb}
\usepackage{amsmath}
\usepackage{cancel}
\usepackage{tcolorbox}

\def\d{{\rm d}}
\def\br{{\bar{r}}}

\usepackage{graphics,appendix,afterpage,makecell} 

\definecolor{oucrimsonred}{rgb}{0.6, 0.0, 0.0}
\definecolor{persianblue}{rgb}{0.11, 0.22, 0.73}
\definecolor{forestgreen}{rgb}{0.13,0.35,0.13}
\definecolor{lightgray}{rgb}{0.83, 0.83, 0.83}
 \hypersetup{colorlinks, citecolor=oucrimsonred, linkcolor=black, urlcolor=oucrimsonred}
\definecolor{cornellred}{rgb}{0.7, 0.11, 0.11}
\definecolor{navyblue}{rgb}{0.0, 0.0, 0.5}
\definecolor{amethyst}{rgb}{0.6, 0.4, 0.8}
\definecolor{yellow}{rgb}{1.0, 1.0, 0.0}
\definecolor{firebrick}{rgb}{0.7, 0.13, 0.13}
\definecolor{tangerineyellow}{rgb}{1.0, 0.8, 0.0}
\definecolor{deepfuchsia}{rgb}{0.76, 0.33, 0.76}
\definecolor{amber}{rgb}{1.0, 0.75, 0.0}
\definecolor{VioletRed4}{rgb}{0.55, 0.13, .32}
\definecolor{indiagreen}{rgb}{0.07, 0.53, 0.03}
\definecolor{VioletRed4}{rgb}{0.55, 0.13, .32}
\newcommand{\be}{\begin{equation}}
\newcommand{\ee}{\end{equation}}
\newcommand{\bea}{\begin{equation} \begin{aligned}}
\newcommand{\eea}{\end{aligned} \end{equation}}

\definecolor{oucrimsonred}{rgb}{0.6, 0.0, 0.0}
\newcommand\vertarrowbox[3][6ex]{%
  \begin{array}[t]{@{}c@{}} #2 \\
  \left\uparrow\vcenter{\hrule height #1}\right.\kern-\nulldelimiterspace\\
  \makebox[0pt]{\scriptsize#3}
  \end{array}%
}

\definecolor{verdechiaro}{rgb}{0.6,1,0.6}
\definecolor{giallochiaro}{rgb}{1,1,0.6}
\definecolor{bluscuro}{rgb}{0.15, 0.2, 0.9}
\definecolor{verdes}{rgb}{0.1, 0.5, 0.1}%
\definecolor{tangerineyellow}{rgb}{1.0, 0.8, 0.0}

\definecolor{americanrose}{rgb}{1.0, 0.01, 0.24}
\definecolor{cobalt}{rgb}{0.0, 0.28, 0.67}
\definecolor{brandeisblue}{rgb}{0.0, 0.44, 1.0}
\definecolor{mycolor}{rgb}{0.0, 0.0, 0.5}
\definecolor{oxfordblue}{rgb}{0.0, 0.13, 0.28}
\definecolor{azure}{rgb}{0.0, 0.5, 1.0}
\definecolor{turquoiseblue}{rgb}{0.0, 1.0, 0.94}
\newtcolorbox{mynewbox}[1]{colback=white!5!white,colframe=azure!75!black,fonttitle=\bfseries,title=#1}
\newtcolorbox{mybox}{colback=mycolor!5!white,colframe=azure!75!black}
\newtcolorbox{mynamedbox}[1]{colback=mycolor!5!white,colframe=azure!75!black,title=#1}
\definecolor{venetianred}{rgb}{0.78, 0.03, 0.08}
\newtcolorbox{mynamedbox1}[1]{colback=venetianred!5!white,colframe=venetianred!80!black,title=#1}
\newtcolorbox{mynamedbox2}[1]{colback=azure!5!white,colframe=azure!80!black,title=#1}

\definecolor{verdes}{rgb}{0.1, 0.5, 0.1}%
\definecolor{cornellred}{rgb}{0.7, 0.11, 0.11}

\definecolor{VioletRed4}{rgb}{0.55, 0.13, .32}

\hypersetup{
     colorlinks   = true,
     citecolor    = violet,
     urlcolor     = violet,
     linkcolor    = violet}

\definecolor{rossocorsa}{rgb}{0.83, 0.0, 0.0}

\usepackage[normalem]{ulem}

\title{
Non-vanishing non-linear
 Static Love Number of a Class of  Extremal Reissner-Nordstr\"om Black Holes }

\author[a]{L.-R. Gounis\orcidlink{0009-0009-6865-2981},}
\author[a]{A. Kehagias\orcidlink{0000-0001-6080-6215},}
\author[a]{G. Panagopoulos\orcidlink{0009-0004-1213-2585},}
\author[b,c]{A. Riotto\orcidlink{0000-0001-6948-0856}}
\affiliation[a]{Physics Division, National Technical University of Athens, Athens, 15780, Greece}
\affiliation[b]{Department of Theoretical Physics, 
24 quai E. Ansermet, CH-1211 Geneva 4, Switzerland}

\affiliation[c]{ Gravitational Wave Science Center,  
24 quai E. Ansermet, CH-1211 Geneva 4, Switzerland}

\abstract{We compute the tidal Love numbers for a particular axially symmetric  configuration of extremal Reissner–Nordström geometry. By exactly solving the non-linear Einstein equations, we investigate the tidal response of extremal Reissner–Nordström black holes in four-dimensional spacetimes under external gravitational fields. We show that, for the specific geometry considered, the static tidal Love number remains finite and non-vanishing to all orders in the external tidal field. By contrast, we verify that the Love number of an isolated extremal Reissner–Nordström black hole remains zero, in agreement with previous expectations. Furthermore, we explicitly calculate the Zerilli–Moncrief master functions and match them with the effective field theory description.

}

\emailAdd{el19081@mail.ntua.gr}
\emailAdd{kehagias@central.ntua.gr}
\emailAdd{gpanagopoulos@mail.ntua.gr}
\emailAdd{antonio.riotto@unige.ch}

\begin{document}
\maketitle

\section{Introduction}
Gravitational waves (GWs) and black holes (BHs) represent cornerstone predictions of general relativity, validated by landmark observations such as the detection of GWs originating from BH coalescences by the LIGO and Virgo collaborations~\cite{LIGOScientific:2021sio}. These observations have provided strong empirical support for Einstein's theory of gravity, with no observed deviations from GR to date~\cite{Kehagias:2024yyp}.

During the inspiral of compact binaries—whether involving neutron stars or BHs—tidal forces grow significant as the orbital separation decreases. These interactions influence both the dynamics of the binary system and the characteristics of the emitted GWs. Accurately modeling this interplay is essential for refining waveform templates and testing GR under highly relativistic conditions.

Tidal interactions are quantified by parameters called Love numbers, which describe how an object deforms under the gravitational field of its companion. In particular, the static tidal Love numbers (TLNs) depend on the internal composition and structure of the compact objects subject to deformation~\cite{poisson_will_2014}. These TLNs first appear at the fifth post-Newtonian order in the GW phase evolution~\cite{Flanagan:2007ix}, and are especially informative for neutron stars, where their nonzero values can reveal properties of dense nuclear matter. 

By contrast, BHs are thought to have zero TLNs due to the absence of material rigidity. This has been demonstrated through linear perturbation theory, which shows that a tidal perturbation proportional to 
$r^\ell$  fails to generate an $r^{-\ell-1}$ response, indicating a vanishing static TLN for each multipole order $\ell$. Consequently, BHs do not develop a measurable static tidal deformation under linear perturbations~\cite{Binnington:2009bb,Damour:2009vw,Damour:2009va,Pani:2015hfa,Pani:2015nua,Porto:2016zng,LeTiec:2020spy,Chia:2020yla,LeTiec:2020bos,Poisson:2021yau,Kehagias:2024yzn}. This outcome is believed to be connected to certain hidden symmetries in the spacetime structure~\cite{Hui:2020xxx,Charalambous:2021mea,Charalambous:2021kcz,Hui:2021vcv,Hui:2022vbh,Charalambous:2022rre,Ivanov:2022qqt,Katagiri:2022vyz,Bonelli:2021uvf,Kehagias:2022ndy,BenAchour:2022uqo,Berens:2022ebl,DeLuca:2023mio,Rai:2024lho}.

Recent studies have further confirmed that the static TLNs remain zero even under second-order perturbations in the external tidal field~\cite{Riva:2023rcm,Riva:2024}. By using the Ernst formalism~\cite{Ernst:1967wx} and Weyl coordinate framework,  the vanishing of the static TLN has been finally proven to persist to all orders in the parity-even tidal deformations in Refs. ~\cite{Kehagias:2024rtz,Combaluzier-Szteinsznaider:2024sgb} for the  the Schwarzschild BH and  in Ref. \cite{Gounis:2024hcm} for the 
rotating Kerr BHs. Furthermore, it has been understood that the vanishing of the  non-linear static TLN is intimately related to the presence of 
  underlying non-linear symmetries. 

In this paper we address the question of whether the non-linear static TLN vanishes for Extremal Reissner-N\"ordstrom (ERN) BHs which generalize  Schwarzschild solutions by adding the electric charge as an extra charge besides the mass. We will show that the ERN BHs are indeed responding to the presence of a tidal force and develop a nonvanishing static TLN. This result is obtained by solving exactly Einstein equations by immersing the ERN BH in a external gravitational field and is therefore valid at any order in the external tidal force. 

The paper is organized as follows. In section 2, we describe the external  spherically symmetric RN solution. In section 3, we analyze a charged axially symmetric extremal RN. For small parameters, it may be regarded as   a perturbation of the standard RN. This is analyzed  in section 4, where the corresponding Zerilli-Moncrief fields and the corresponding Love numbers  are found. In section 5 we match with the EFT   and, finally,  in section 6 we conclude.

\section{The Extremal RN BH}

Among the most elegant and instructive exact solutions of Einstein's field equations is the Reissner--Nordstr\"om (RN) BH. It represents a non-rotating, spherically symmetric BH that possesses electric charge, and serves as a natural generalization of the Schwarzschild solution. Whereas the Schwarzschild BH is fully characterized by its mass alone, the RN solution is described by two parameters: the mass \( M \) and the electric charge \( Q \). This added degree of freedom significantly enriches the geometry and causal structure of the spacetime. QNM of the RN BH have been studied in \cite{Kokkotas:1988fm,Berti:2003ud,Berti:2003zu}.

The RN solution emerges from Einstein's equations coupled to the source-free Maxwell equations. It captures the geometry of spacetime outside a charged, non-rotating, spherically symmetric object. The presence of the electromagnetic field modifies the gravitational dynamics in a subtle but profound way. Most notably, the RN spacetime admits, in general, two distinct horizons: an outer event horizon and an inner Cauchy horizon. These arise provided the charge is smaller than the mass (\( |Q| < M \)). In the extremal case (\( |Q| = M \)), the two horizons coincide, leading to a degenerate horizon with special thermodynamic properties. If the charge exceeds the mass (\( |Q| > M \)), the solution no longer describes a BH, but a naked singularity—raising questions about cosmic censorship.

The RN BH stands out not just for its mathematical elegance, but also for its physical implications. It offers a useful setting for investigating a range of theoretical problems, such as the stability of inner horizons, the behavior of test particles in charged spacetimes, and the nature of singularities. Furthermore, the extremal Reissner--Nordstr\"om BH plays a pivotal role in studies of BH thermodynamics and quantum gravity, as it features zero surface gravity and thus vanishing Hawking temperature. These properties make it an important model for understanding extremality and horizon microstructure in various approaches to quantum gravity.


Let us consider the gravitational field of a charged object with mass \( M \) and electric charge \( Q \), assuming a static, spherically symmetric spacetime. The corresponding line element takes the form
\begin{equation}
    \d s^2 = -f(r)\, \d t^2 +\frac{1}{f(r)} \, \d r^2 + r^2(\d\theta^2 + \sin^2\theta\, \d\phi^2) 
\end{equation}
where the \( f(r) \) is given by 
\begin{equation}
    f(r)= 1 - \frac{2M}{r} + \frac{Q^2}{r^2}.
    \label{fr}
\end{equation}
The associated electrostatic potential is given by
\begin{equation}
    A_\mu =  \Phi(r)\, \delta_\mu^0, \qquad \Phi(r) = \frac{Q}{r}. \label{ff}
\end{equation}
There is a singularity at $r=0$, which is hidden behind the two horizons located at 
\begin{equation}
    r_\pm=M\pm \sqrt{M^2-Q^2}.
\end{equation}
Clearly, the bound 
\begin{equation}
    M\geq |Q|, \label{Mq}
\end{equation}
should be satisfied, since otherwise, there is no horizon and  $r=0$ is a naked singularity, violating the cosmic censorship hypothesis. When the bound (\ref{Mq}) is saturated, i.e., 
\begin{equation}
    M=|Q|, \label{mq}
\end{equation}
the two horizons 
coincide with $r_\pm=M$. This is the  particular case of the RN spacetime,  the Extremal RN  (ERN) spacetime with metric 
\begin{equation}
    \d s^2 = -\left(1-\frac{M}{r}\right)^2\, \d t^2 +\dfrac{1}{\left(1-\dfrac{M}{r}\right)^2} \, \d r^2 + r^2(\d\theta^2 + \sin^2\theta\, \d\phi^2).
    \label{mern}
\end{equation}
The ERN has a single horizon at $r_s=M$, and its near horizon geometry ($r\approx M+\rho$) turns out to be 
\begin{equation}
    \d s^2 = -\frac{\rho^2}{M^2}\, \d t^2 +\frac{M^2}{\rho^2} \, \d \rho^2 + M^2(\d\theta^2 + \sin^2\theta\, \d\phi^2).
    \label{NHG}
\end{equation}
This is the metric of AdS$_2$$\times$S$^2$, so that the near horizon geometry is a smooth manifold. 
It is useful to make the change of coordinates 
\begin{equation}
    \br=r-M, \label{rbar}
\end{equation}
so that the metric (\ref{mern}) is written as 
\begin{equation}
    \d s^2 = -\dfrac{1}{\left(1+\dfrac{M}{\br}\right)^2}\, \d t^2 +\left(1+\frac{M}{\br}\right)^2 \, \bigg[\d \br^2 + \br^2(\d\theta^2 + \sin^2\theta\, \d\phi^2)\bigg],
    \label{mern1}
\end{equation}
whereas, the horizon now is at 
\begin{equation}
    \br=0.
\end{equation}
Similarly, the electrostatic potential turns out to be 
\begin{equation}
    \Phi(\br) = \frac{M}{\br +M}. \label{phin}
\end{equation}

\section{The Axisymmetric Extremal RN BH}

 The metric in Eq. (\ref{mern1}) can be generalized  to the general form \cite{papapetrou}
\begin{equation}
    \d s^2 = -\dfrac{1}{\left(1+\psi\right)^2}\, \d t^2 +\left(1+\psi\right)^2 \, \bigg[\d \br^2 + \br^2(\d\theta^2 + \sin^2\theta\, \d\phi^2)\bigg] .
    \label{mern2}
\end{equation}
and
\begin{equation}
    \Phi=\frac{\psi}{1+\psi}, \label{PP}
\end{equation}
where $\psi=\psi(\vec{r})$. 
 Einstein equations are satisfied if 
\begin{equation}
    \frac{\nabla^2\psi}{(1+\psi)^5}=0.
\end{equation}
In other words,  if $\psi$ obeys the Laplace equation 
\begin{equation}
    \nabla^2\psi=0, \label{psi}
\end{equation}
on the flat three-dimensional Euclidean space, then  Einstein equations are automatically satisfied. In addition,  Maxwell equations are written as 
\begin{equation}
\vec{\nabla}\left[(1+\psi)^2\vec{\nabla} \left(\frac{\psi}{1+\psi}\right)\right]=0,
\end{equation}
and are  automatically satisfied once  Eq. (\ref{psi}) holds. 

The spherically symmetric ERN corresponds
to the spherically symmetric solution 
\begin{equation}
    \psi(\br)=\frac{M}{\br}. \label{psio}
\end{equation}
Clearly, there are more general solutions to Eq. (\ref{psi}) which are not spherically symmetric but rather axially symmetric. We will refer to such  solutions of (\ref{psi})  
as  Axisymmetric Extremal RN (AERN).
As we will see, they describe  ERN  embedded in external gravitational environments and therefore they are useful to study the static TLNs. Such  solutions are of the generic  form
\begin{equation}
    \psi=\frac{M}{\br }+ \sum_{\ell=1} \left(C_\ell \br ^\ell +
    \frac{B_\ell}{\br^{\ell+1}}\right)Y_{\ell m}(\theta,\phi).
    \label{pps}
\end{equation}
Let us notice however, that the metric of the AERN (\ref{mern2}) is singular at the points where  $\psi$ satisfies 
\be
1+\psi=0,\label{1p}
\ee
so that, care should be taken such that 
$\psi$ given in  (\ref{pps}) does not violate (\ref{1p}). 
For the $\ell=2$ case we are interested in, we can take  for $\psi$ the following
\begin{align}
    \psi=&\frac{M}{\br}+\frac{Q}{\sqrt{a^2+\br^2+2 \,a \,\br \cos\theta}}+ \frac{Q}{\sqrt{a^2+\br^2-2\, a \, \br \cos\theta}}-\frac{2Q}{a}\nonumber\\
    &\hspace{0.7cm}+ \frac{Q}{\sqrt{b^2+\br^2+2 \,b \,\br \cos\theta}}+ \frac{Q}{\sqrt{b^2+\br^2-2\, b \, \br \cos\theta}}-\frac{2Q}{r}.
    \label{psif}
\end{align}

This describes the solution  for an extremal RN BHs residing at $\br=0$
together with four other extremal RN BHs sitting at $\theta=0$ and 
\begin{equation}
    z= a,\ -a, \ b \ \ \mbox{and}\ \  -b.
\end{equation}
\setlength{\columnsep}{20pt}
\begin{wrapfigure}{r}{0.24\textwidth}
  \begin{center}
  \vspace{-25pt} 
\includegraphics[width=0.15
\textwidth]{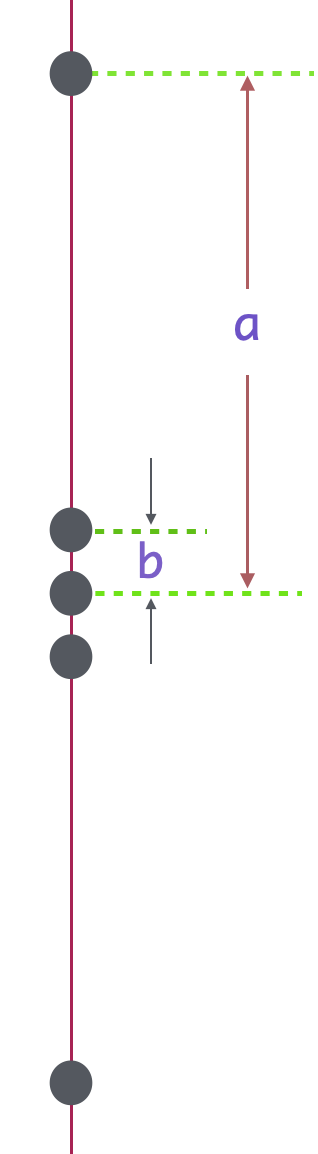}
  \end{center}
  \vspace{-5pt} 
  \caption{The BH configuration of Eq. (\ref{psif}).}
\end{wrapfigure}

\vspace{-30pt}
The one at $\br=0$ is  the central BH and the others are accompanying BHs.
 Clearly, $\psi$ is positive, and hence the metric (\ref{mern2})  is regular everywhere. Eq. (\ref{psif}) is written  equivalently as  
 \begin{equation}
     \psi(\br, \theta)=\frac{M}{r}
     +\frac{Q}{a}\sum_{\ell=1}^\infty P_{2\ell}(\cos\theta)\left(\frac{\br}{a}\right)^{2\ell}+
     \frac{Q}{\br}\sum_{\ell=1}^\infty P_{2\ell}(\cos\theta)\left(\frac{b}{\br}\right)^{2\ell}. \label{leg}
 \end{equation}
It is easy then to see that  in the limit  
\be
a\to \infty,\qquad  b\to 0,
\label{ab}
\ee
such that 
\begin{align}
   \frac{Q}{a^3}&=C_2=\mbox{finite},\nonumber \\  Q\,  b^2&=B_2=\mbox{finite},
   \label{finite}
\end{align}
we get that $\psi$ reduces to
\begin{equation}
    \psi=\frac{M}{\br}+\left(C_2\br^2+\frac{B_2}{\br^3}\right) P_2(\cos \theta).
    \label{psi2}
\end{equation}

The situation is completely analogous of a  conducting sphere in an external constant electric field, where the later is generated by  similarly by an appropriate limit   of two charges at a distance, as in Eqs. (\ref{ab}) and (\ref{finite}). 
In what follows we regard Eq. (\ref{psi2}) as a regulated limit of a smooth geometry that realizes the configuration (\ref{psi2}), so no singularity issue arises. Furthermore, (\ref{psi2}) should not be interpreted as an isolated RN black hole. It is the appropriate scaling limit of a system consisting of two widely separated black holes together with a central cluster of black holes.  


\section{The AERN as tidal perturbation of ERN}

For a small  enough $C_\ell,\,B_\ell
$ so that $C_\ell\ll r/(r-M)^{\ell+1}$ and $\,B_\ell\ll r(r-M)^{\ell}$, the metric (\ref{mern2}) becomes 
\begin{equation}
    \begin{split}
        \rm{d}s^2=&-\left(1-\frac{M}{r}\right)^2\left[1-2\left(\frac{B_\ell}{r^{\ell+1}\left(1-\frac{M}{r}\right)^\ell}+C_\ell r^\ell\left(1-\frac{M}{r}\right)^{\ell+1}\right)\rm{P}_\ell(\cos\theta)\right]\rm{d}t^2 \\ &+\left[1+2\left(\frac{B_\ell}{r^{\ell+1}\left(1-\frac{M}{r}\right)^\ell}+C_\ell r^\ell\left(1-\frac{M}{r}\right)^{\ell+1}\right)\rm{P}_\ell(\cos\theta)\right]\left[\frac{\rm{d}r^2}{\left(1-\frac{M}{r}\right)^2}+r^2\rm{d}\Omega^2\right]
    \end{split}
    \label{metper}
\end{equation}
Similarly, the electrostatic potential becomes:
\begin{equation}
    \Phi=\frac{\psi}{1+\psi}=\underbrace{\frac{\psi_0}{1+\psi_0}}_{\Phi_0}+\frac{\delta\psi}{(1+\psi_0)^2}=\Phi_0+\left(\frac{B_\ell}{r^{\ell+1}\left(1-\frac{M}{r}\right)^{\ell-2}}+C_\ell r^\ell\left(1-\frac{M}{r}\right)^{\ell+3}\right)\rm{P}_\ell(\cos\theta).
\end{equation}
It can be interpreted as a tidal perturbation of the Papapetrou metric.
Notice that the tidal field formally diverges at large distances, where perturbation theory breaks down.  In realistic situations, the external tidal field behaves as an increasing source only within a limited spatial region, and it naturally vanishes as one moves toward spatial infinity. Therefore, assuming that the tidal field proportional to $C_\ell$  is a small perturbation to the dominant $M/r$ term, there is no singularity issue. This means that for $C_\ell=\mathcal{C}_\ell M^{-\ell}$, we have $r\ll M/\mathcal{C}_\ell^{1/\ell}$ 
and the $\psi$ can be regarded as an approximation of an exact solution as we have already noticed.

However, one should not look simply at the metric but instead at the Zerilli-Moncrief function, which we will construct in a moment.
Employing now  the usual parametrization of $\delta g_{\mu\nu}$
as 
\begin{equation}
    \delta g_{\mu\nu}=\bigg(f\, H_0,\frac{H_2}{f},r^2 K,r^2 \sin^2\theta K\bigg) \, P_\ell(\cos\theta),
    \label{metperparametr}
\end{equation}
we can read off the metric perturbations of the ERN from Eq. (\ref{metper}) as 
\begin{equation}
    H_0=H_2=K=2\left(\frac{B_\ell}{r^{\ell+1}\left(1-\frac{M}{r}\right)^\ell}+C_\ell r^\ell\left(1-\frac{M}{r}\right)^{\ell+1}\right). 
\end{equation}
Similarly, the perturbation of the electrostatic potential $\delta\Phi$ turns out to be
\begin{equation}
    \delta \Phi= \left[C_\ell \, r^\ell \left(1-\frac{M}{r}\right)^{\ell+2} + \frac{B_\ell}{r^{\ell+1}\left(1-\frac{M}{r}\right)^{\ell-1}} \right]P_\ell(\cos\theta). 
\end{equation} 
 The electric field  is
\begin{equation}
     E^i=-\sqrt{-g}\, F^{i0},
\end{equation}
so that the perturbation of the electrostatic potential $\delta \Phi$ gives rise to a corresponding perturbation of the  radial electric field  
\begin{equation}
    \delta E^r=E \sin\theta P_\ell (\cos\theta), 
\end{equation}
where $E$ is given by
\begin{equation}
    E=C_\ell \, \ell (r-M)^{\ell+1}-\frac{\ell(\ell+1) B_\ell}{(r-M)^\ell}.
    \label{M}
\end{equation}
We can  now define the quantities \cite{Moncrief:1974ng, Chandrasekhar:1979iz,Xanthopoulos} 
\begin{align}
    Z_1=&\frac{Q_1}{\Lambda}\sqrt{(l-1)(l+2)}, \nonumber \\
    Z_2=&-2 E-\frac{2M}{r} \frac{Q_1}{\Lambda},
    \label{HH}
\end{align}
    where 
    \begin{equation}
        Q_1=2r e^{-2\lambda}\left[ H_2-(1+r \partial_r\lambda)K-r \partial_r K\right]+\ell(\ell+1)\,r\,K, \qquad e^{\lambda}=\left(1-\frac{M}{r}\right)^{-1},
    \end{equation}
and 
\begin{equation}
    \Lambda=(\ell-1)(\ell+2)+\frac{6M}{r}-\frac{4M^2}{r^2}.
\end{equation}
Then, the Zerilli-Moncrief master functions \cite{Zerilli:1970se,Moncrief:1974ng} are 
\begin{align}
    \Psi_1=& \sqrt{\frac{\ell-1}{2\ell+1}} Z_1
    -\sqrt{\frac{\ell+2}{2\ell+1}}Z_2,
    \nonumber \\
    \Psi_2=&\sqrt{\frac{\ell+2}{2\ell+1}}Z_1+
    \sqrt{\frac{\ell-1}{2\ell+1}}Z_2.
    \label{Psi12}
\end{align}
After calculations, these turn out to be
\begin{align}
    \Psi_1=& \frac{2 C_\ell\, \ell\sqrt{(2\ell+1)(\ell+2)}}{(\ell+2) r-2M} \, \, r^{\ell+2} \left(1-\frac{M}{r}\right)^{\ell+1}
   , \label{P1}\\
    \Psi_2=&\frac{2 B_\ell\, (\ell+1) \sqrt{(2\ell+1)(\ell-1)}}{(\ell-1) r+2M} \, \, r^{-\ell+1} \left(1-\frac{M}{r}\right)^{-\ell}. \label{P2}
\end{align}
They should satisfy the (time-independent)  Zerilli-Moncrief equation \cite{Moncrief:1974ng}
\begin{equation}
    -\frac{{\rm d}^2}{{\rm d}r_*^2}\Psi_i+V_i \Psi_i=0, \qquad i=1,2, \label{pss}
\end{equation}
where, for the extreme RN, 
\begin{equation}
    \frac{{\rm d}r}{{\rm d}r_*}=\left(1-\frac{M}{r}\right)^2,
\end{equation}
and
\begin{align}
V_1=&\frac{(r-M)^2}{r^6\Big[(\ell+2) r-2M\Big]^2}\bigg[\ell(\ell+1)(\ell+2)^2 r^4-2(\ell-1)(\ell+2)^2M r^3\nonumber \\
   & \hspace{5cm}-12 (\ell+2) M^2 r^2+8(\ell+3)M^3r-8 M^4
    \bigg], \label{V1}
\\ \nonumber 
\\
    V_2=&\frac{(r-M)^2}{r^6\Big[(\ell-1) r+2M\Big]^2}\bigg[-2 \left(l^3-3 l+2\right) M r^3+8 (l-2) M^3 r\nonumber \\
   & \hspace{5cm}-12 (l-1) M^2 r^2-(l-1)^2 l (l+1) r^4+8 M^4\bigg]. \label{V2}
\end{align}
An equivalent expression for the Zerilli-Moncrief equation is
\begin{equation}
    -\left(1-\frac{M}{r}\right)^2\frac{\rm{d}}{\rm{dr}}\left(\left(1-\frac{M}{r}\right)^2\frac{\rm{d}\Psi_i}{\rm{d}r}\right)+V_i \Psi_i=0, \qquad i=1,2, \label{pss2}
\end{equation}
Defining now the static TLNs $k^\ell$ as the coefficients in the expansion for $r\gg M$ of the linear combination
\begin{equation}
    a_\ell \Psi_{1} + b_\ell \Psi_{2} \sim r^{\ell+1}
    \left[1+\cdots+k^\ell\left(\frac{1}{r}\right)^{2\ell+1}+\cdots\right],
\end{equation}
for some (to be specified) coefficients $a_\ell, b_\ell$, we obtain
\begin{equation}
    a_\ell \Psi_{1} + b_\ell \Psi_{2} = 2 a_\ell C_\ell \ell \sqrt{\frac{2\ell+1}{\ell+2}} r^{\ell+1}\left[1 + \frac{b_\ell}{a_\ell} \frac{B_\ell}{C_\ell} \left(\frac{\ell+1}{\ell}\right) \sqrt{\frac{\ell+2}{\ell-1}} r^{-2\ell-1} \right]
\end{equation}
and thus the TLN is
\begin{equation}
    k^\ell = \frac{b_\ell}{a_\ell} \frac{B_\ell}{C_\ell} \left(\frac{\ell+1}{\ell}\right) \sqrt{\frac{\ell+2}{\ell-1}}.
\label{TLNZerilliMoncrief}
\end{equation}

Let us stress at this point that for an isolated RN black hole in an external gravitational field we have $C_\ell\neq 0$ and $B_\ell=0$. Indeed, for the $\ell=2$ tidal perturbation described by  $\psi$ of Eq. (\ref{psi2}), we have that  $Qb^2\to 0$ as $b\to 0$, so that the second line in Eq. (\ref{psi2}) is missing. In this case, the charge $Q$ sources an external tidal gravitational field proportional to 
$C_2$ in the particular limit we considered above.  However, because the RN black hole is isolated now, we also have $B_2=Q b^2=0$.
 Thus, we once again confirm that the static tidal Love number of an isolated extremal RN black hole vanishes, i.e., $k^\ell=0$.

\section{Matching with the EFT}
Any object, observed from sufficiently large distances, appears as a point source. Therefore, we may make use of the point-particle effective field theory (EFT) in order to describe the properties of a finite-size object from the point of view of an observed located far away from it. In the EFT approach the tidal Love numbers are defined without ambiguities, which in the case of General Relativity are caused by non-linearities and gauge invariance. More specifically, if we consider the point-particle EFT of a massive, charged, self-gravitating object, the EFT action can be written as \cite{Rai:2024lho}
\begin{equation}
    S_{EFT} = S_{\text{bulk}} + S_{\text{pp}} + S_{\text{finite-size}}
\end{equation}
where
\begin{equation}
    S_{\text{bulk}} = \frac{1}{16\pi} \int \d^4 x \, \sqrt{-g} (R-F_{\mu \nu} F^{\mu \nu})
\end{equation}
is the Einstein-Maxwell action describing gravitation and electromagnetism in the bulk,
\begin{equation}
    S_{\text{pp}} = \int \d\tau \left(-M\sqrt{g_{\mu \nu} \frac{dx^\mu}{d\tau}\frac{dx^\nu}{d\tau}} + Q A_\mu \frac{dx^\mu}{d\tau}\right)
\end{equation}
describes a charged point-particle, and
\begin{equation}
    \begin{split}
    S_{\rm finite-size}  = \sum_{\ell=1}^{\infty}  \int \d \tau & \Bigg[ \frac{\lambda_{\ell}^{(E)}}{2\ell!}\left(\partial_{(a_{1}}\cdots\partial_{a_{\ell-1}}E_{a_{\ell})_{T}}\right)^{2} + \frac{\lambda_{\ell}^{(B)}}{4\ell!}\left(\partial_{(a_{1}}\cdots\partial_{a_{\ell-1}}B_{a_{\ell})_{T}b}\right)^{2}\\ 
    & + \frac{\lambda_{\ell}^{(C_{E})}}{2\ell!} \left(\partial_{(a_{1}}\cdots\partial_{a_{\ell-2}}E_{a_{\ell-1}a_{\ell})_{T}}^{(2)}\right)^{2} +\frac{\lambda_{\ell}^{(C_{B})}}{4\ell!}\left(\partial_{(a_{1}}\cdots\partial_{a_{\ell-2}}B_{a_{\ell-1}a_{\ell})_{T}|b}^{(2)}\right)^{2} \\ 
    & +\frac{\eta_{\ell}^{(E)}}{\ell!}\left(\partial_{(a_{1}}\cdots\partial_{a_{\ell-1}}E_{a_{\ell})_{T}}\right)\left(\partial_{(a_{1}}\cdots\partial_{a_{\ell-2}}E_{a_{\ell-1}a_{\ell})_{T}}^{(2)}\right)\\ 
    & +\frac{\eta_{\ell}^{(B)}}{2\ell!}\left(\partial_{(a_{1}}\cdots\partial_{a_{\ell-1}}B_{a_{\ell})_{T}b}\right)\left(\partial_{(a_{1}}\cdots\partial_{a_{\ell-2}}B_{a_{\ell-1}a_{\ell})_{T}|b}^{(2)}\right)\Bigg]. 
    \end{split}
    \label{eq:Sfinitesize}
\end{equation}
Here, $(\cdots)_T$ denotes the symmetrized traceless component of the enclosed indices, $E_a$ and $B_a$ are the electric and magnetic field
\begin{equation}
    E_a \equiv F_{0a} = \dot{A}_a - \partial_a A_0 \, , \qquad B_{ab} \equiv F_{ab} = \partial_a A_b - \partial_b A_a \, ,
\end{equation}
and $E_{ab}^{(2)}, B_{ab}^{(2)}$ are the electric and magnetic components of the Weyl tensor,
\begin{equation}
    E^{(2)}_{ab} \equiv C_{0a0b} = - \frac{1}{2}\partial_a\partial_b h_{00} \, , \qquad B^{(2)}_{ab\vert c} \equiv C_{0abc} =  \frac{1}{2}\left( \partial_a \partial_{b} h_{c0} -\partial_a \partial_{c} h_{b0} \right)
    =\partial_a \partial_{[b} h_{c]0} \,.
\end{equation}
The tidal Love numbers are therefore defined to be the coefficients of the effective field theory couplings. There are pure electromagnetic $(\lambda_\ell^{(E)}, \lambda_\ell^{(B)})$, pure gravitational $(\lambda_\ell^{(C_E)}, \lambda_\ell^{(C_B)})$ and mixed gravitational-electromagnetic $(\eta_\ell^{(E)}, \eta_\ell^{(B)})$ Love numbers. Focusing in our case, where the magnetic parts are absent, calculations in the EFT model give the following expressions for the radial electric field and the linearized Weyl tensor:
\begin{align} 
E_{r}
 & \propto  r^{\ell-1} P_\ell (\cos{\theta}) \Bigg[ \bar{d}^{(E)} \left(  1 -(-1)^\ell \frac{2^{\ell}\sqrt{\pi}(\ell+1)}{\ell \, \Gamma(\frac{1}{2}-\ell)} \lambda_\ell^{(E)} r^{-2\ell-1} \right) \nonumber \\& \quad\quad\quad\quad\quad\quad\quad\quad\quad\quad\quad\quad\quad\quad\quad\quad
 -\bar{c}^{(E)}(-1)^\ell \frac{2^{\ell-1}\sqrt{\pi}(\ell+1)}{\ell \, \Gamma(\frac{1}{2}-\ell)}    \eta^{(E)}_\ell r^{-2\ell-1} \Bigg] \, , \label{ErEFTSantoni}
\\
 C_{0r0r}
& \propto r^{\ell-2} P_\ell (\cos{\theta})  \Bigg[  \bar{c}^{(E)} \left( 1 +  (-1)^\ell  \frac{(\ell+1)(\ell+2)}{\ell(\ell-1)}\frac{2^{\ell}\sqrt{\pi}}{\Gamma(\frac{1}{2}-\ell)}  \lambda_\ell^{(C_E)} r^{-2\ell-1}   \right)
\nonumber \\ 
&\qquad\qquad\qquad\qquad\qquad\qquad\ 
  +  \bar{d}^{(E)}(-1)^\ell  \frac{(\ell+1)(\ell+2)}{\ell(\ell-1)}\frac{2^{\ell+1}\sqrt{\pi}}{\Gamma(\frac{1}{2}-\ell)}  \eta^{(E)}_\ell  r^{-2\ell-1}   \Bigg] \, . \label{C0r0rEFTSantoni}
\end{align}
Given a metric perturbation of the form (\ref{metperparametr}) and a electrostatic potential perturbation $\delta \Phi$, the expressions for the radial electric and linearized Weyl tensor are given by
\begin{equation}
    E_{r} = - \partial_r (\delta \Phi),
\end{equation}
\begin{equation}
    C_{0r0r} = - \frac{1}{2} \partial^2_r \left(\frac{\Delta}{r^2} H_0 (r)\right) + \frac{2 M}{r^2} \partial_r (\delta \Phi),
\end{equation}
where $\Delta=(r-M)^2$ for the extremal RN black hole. If we calculate these, we find that
\begin{align}
    E_r =& r^{-\ell-3}\left(1-\frac{M}{r}\right)^{-\ell} \left[\left((\ell+1)r - 2M\right) B_\ell - r^{2\ell + 1}\left(1-\frac{M}{r}\right)^{2\ell+1} (\ell r +2M) C_\ell \right], \\
    C_{0r0r} =& - r^{-\ell-5}\left(1-\frac{M}{r}\right)^{-\ell} \Bigg[\left((\ell+1)(\ell+2)r^2 - 2 (2\ell + 5) M r + 8 M^2 \right) B_\ell \nonumber \\ &   \ \ \qquad\qquad\qquad
    + r^{2\ell+1}\left(1-\frac{M}{r}\right)^{2\ell+1}(\ell (\ell-1) r^2 + 2(2\ell-3)M r + 8 M^2) C_\ell\Bigg].   
\end{align}
After a few computations, we can find that in the limit $M \ll r$ we have
\begin{align}
    E_r =& r^{\ell-1} P_\ell (\cos{\theta}) \Bigg[(\ell+1) B_\ell \sum_{n=0}^\infty \binom{\ell+n-1}{n} \left(\frac{M}{r}\right)^n r^{-2\ell-1} \nonumber\\ & \qquad \qquad \qquad \qquad \qquad \qquad \qquad \qquad \qquad - \ell C_\ell \sum_{k=0}^{\ell+1} \binom{\ell+1}{k} (-1)^k \left(\frac{M}{r}\right)^k \Bigg]
\end{align}
Since $M \ll r$ we can approximately keep only the $n=k=0$ terms and get
\begin{equation}
    E_r \simeq -\ell C_\ell r^{\ell-1} P_\ell (\cos{\theta}) \left[1 - \frac{(\ell+1)}{l} \frac{B_\ell}{C_\ell} r^{-2\ell-1}\right].
    \label{ErEFTresult}
\end{equation}
A similar process for $C_{0r0r}$ yields
\begin{equation}
    C_{0r0r} \simeq -(\ell+1) \ell C_\ell r^{\ell-2} P_\ell (\cos{\theta}) \left[1 + \frac{(\ell+2)}{\ell} \frac{B_\ell}{C_\ell} r^{-2\ell-1} \right].
    \label{C0r0rEFTresult}
\end{equation}
These expressions should be compared  with eqs. (\ref{ErEFTSantoni}) and (\ref{C0r0rEFTSantoni}) in order to extract the TLNs. A quick glance makes it obvious that, while we have four matching conditions, we have five unknowns: $\bar{c}^{(E)},\bar{d}^{(E)},\lambda^{(E)}_\ell, \lambda^{(C_E)}_\ell, \eta^{(E)}_\ell$. However, after some algebra, we can get
\begin{align}
    E_r &\propto \bar{d}^{(E)} r^{\ell-1} P_\ell (\cos{\theta}) \left[1+\left(\lambda^{(E)}_\ell + \frac{1}{2} \frac{\bar{c}^{(E)}}{\bar{d}^{(E)}} \eta^{(E)}_\ell \right) f_\ell \, r^{-2\ell-1}\right], \\
    C_{0r0r} &\propto \bar{c}^{(E)} r^{\ell-2} P_\ell (\cos{\theta}) \left[1 + \left(\lambda^{(C_E)} + 2 \frac{\bar{d}^{(E)}}{\bar{c}^{(E)}} \eta^{(E)}_\ell \right) q_\ell \, r^{-2\ell-1}\right],
\end{align}
where
\begin{align}
    f_\ell &= (-1)^{\ell+1} \frac{ (\ell+1) 2^\ell \sqrt{\pi}}{\ell \Gamma\left(\frac{1}{2}-\ell\right)}, \\
    q_\ell &= (-1)^{\ell} \left(\frac{\ell+2}{\ell-1}\right) \frac{ (\ell+1) 2^\ell \sqrt{\pi}}{\ell \Gamma\left(\frac{1}{2}-\ell\right)} = \left(\frac{\ell+2}{\ell-1}\right) f_\ell.
\end{align}
If we now define
\begin{align}
    \bar{\lambda}^{(E)}_\ell&=\left(\lambda^{(E)}_\ell+\frac{1}{2} \frac{\bar{c}^{(E)}}{\bar{d}^{(E)}} \eta^{(E)}_\ell\right) f_\ell, \\
\bar{\lambda}^{(C_E)}_\ell &= \left(\lambda^{(C_E)}_\ell + 2 \frac{\bar{d}^{(E)}}{\bar{c}^{(E)}} \eta^{(E)}_\ell\right) q_\ell,
\end{align}
we obtain
\begin{align}
    E_r & \propto \bar{d}^{(E)} r^{\ell-1} P_\ell (\cos{\theta}) \left[1+ \bar{\lambda}^{(E)}_\ell r^{-2\ell-1}\right], \\
    C_{0r0r} &\propto \bar{c}^{(E)} r^{\ell-2} P_\ell (\cos{\theta}) \left[1 + \bar{\lambda}^{(C_E)}_\ell r^{-2\ell-1}\right].
\end{align}
Therefore, while it is impossible to determine $\lambda^{(E)}_\ell, \lambda^{(C_E)}_\ell, \eta^{(E)}_\ell$ by matching, it is possible to determine $\bar{\lambda}^{(E)}_\ell$ and $\bar{\lambda}^{(C_E)}_\ell$, i.e. the "shifted" TLNs by the gravitational-electromagnetic mixing. Matching with our expressions (\ref{ErEFTresult}) \& (\ref{C0r0rEFTresult}) yields
\begin{align}
    \bar{\lambda}^{(E)}_\ell &= - \frac{(\ell+1)}{\ell} \frac{B_\ell}{C_\ell}, \\
    \bar{\lambda}^{(C_E)}_\ell &= \frac{(\ell+2)}{\ell} \frac{B_\ell}{C_\ell} .\label{TLNEFTresult}
\end{align}
Now, all that is left is to equate the two expressions (\ref{TLNZerilliMoncrief}) \& (\ref{TLNEFTresult}) (i.e. $k^\ell = \bar{\lambda}^{(C_E)}_\ell$) for the TLN in order to find the appropriate linear combination of $\Psi_1, \Psi_2$. We find that
\begin{equation}
    b_\ell = \frac{\sqrt{(\ell-1)(\ell+2)}}{\ell +1} a_\ell.
\end{equation}
Therefore, the appropriate linear combination is
\begin{equation}
\label{a}
    \Psi = \Psi_1 + \frac{\sqrt{(\ell-1)(\ell+2)}}{\ell +1} \Psi_2.
\end{equation}

\section{Conclusions}

 Our construction isolates two irreducible pieces of the static, axisymmetric response of an ERN BH embedded in a distant,  tidal environment. The coefficients \(C_{\ell}\) and \(B_{\ell}\) play conjugate roles: \(C_{\ell}\) multiplies the growing, far–zone solution that represents the applied tidal field of the remote companions, while \(B_{\ell}\) multiplies the decaying, near–zone solution that encodes the induced multipole. In gauge–invariant language this separation is realized by the Zerilli–Moncrief master fields, with \(\Psi_{1}\propto C_{\ell}\) and \(\Psi_{2}\propto B_{\ell}\). Horizon regularity together with asymptotic matching selects a unique observable combination (\ref{a})
whose far–zone behavior determines the measurable response. The entire static polarizability is then governed by one dimensionless ratio:
\begin{equation}
k^{\ell} \;=\; \bar{\lambda}^{(C_E)}_{\ell} \;=\; \frac{\ell+2}{\ell}\,\frac{B_{\ell}}{C_{\ell}}, \qquad (r\gg M)\,,
\end{equation}
in exact agreement between the wave–mechanics (Zerilli–Moncrief) and point–particle EFT descriptions derived in the main text.

In addition, within a transparent perturbative regime (small \(B_{\ell},C_{\ell}\) and \(r\ll M/\mathcal{C}_{\ell}^{1/\ell}\)), we established a one–parameter description of the static polarizability of an ERN black hole under a geometrically realized, multi–source tide. Isolated black holes retain vanishing static TLNs.  Tidally coupled, multi–source configurations behave as polarizable objects with an effective response encoded in \(B_{\ell}/C_{\ell}\) and captured identically by the Zerilli–EFT matching.

\normalsize
\begin{acknowledgments}
\noindent
We would like to thank P. Pani and L. Hui for correspondence and L. Santoni for reading the manuscript.
     A.R.  acknowledges support from the  Swiss National Science Foundation (project number CRSII5\_213497)
and by  the Boninchi Foundation for the project ``PBHs in the Era of GW Astronomy''.
\end{acknowledgments}

\newpage

\bibliographystyle{JHEP}
\bibliography{ERN}
\end{document}